# Microwave Sintering of Alumina at 915 MHz: Modeling, Process Control and Microstructure Distribution

**Sylvain Marinel [1,*], Charles Manière [1] , Anthony Bilot [1], Christelle Bilot [1], Christelle Harnois [1], Guillaume Riquet [1], François Valdivieso [2], Christophe Meunier [2], Christophe Coureau [3], François Barthélemy [4]**

[1] Laboratoire de Cristallographie et Sciences des Matériaux, Normandie Univ, ENSICAEN, UNICAEN, CNRS, CRISMAT, 14000 CAEN, France

[2] Laboratoire Georges Friedel, École des Mines de Saint-Étienne, 158, cours Fauriel, CS 62362 F-42023 SAINT-ÉTIENNE cedex 2, France

[3] SOLCERA, ZI n°1 Rue de l''industrie, 27000 EVREUX

[4] DGA (Direction Générale de l'Armement), Echangeur de Guerry, 18000 BOURGES

* Corresponding author: sylvain.marinel@ensicaen.fr, tel: +33 (0)2 31 45 13 69 (France)

**Abstract:** Microwave energy can be advantageously used for materials processing as it provides high heating rates and homogeneous temperature field distribution. These features are partly due to the large microwave penetration depth into dielectric materials which is, at room temperature, a few centimeters in most dielectric materials. However, up to now, this technology is not widely spread for high-temperature materials processing applications (>1200°C), because its reproducibly and ability to sinter large size samples (>30 cm³) still needs to be improved. In this context, this paper describes both an empirically designed 915 MHz single-mode cavity made from SiC susceptors and refractory thermal insulation, and the 3 D modeling of the process in order to improve our understanding of it. Different susceptors geometries and coupling slit position were numerically tested in order to better understand how these parameters impact the field homogeneity and the process stability. It was found that positioning the largest surface of the susceptors parallel to the electrical field allows a very uniform and hybrid heating of the material, while avoiding plasma or thermal instabilities. This was correlated to the 3 D modeling results. Finally, thanks to a fully automatized system this apparatus was used to sinter large size (~30 cm³) low-loss dielectric alumina samples. The sintered materials were subsequently characterized in terms of density, grains size distribution and homogeneity. The reproducibility was also discussed demonstrating the process efficiency and reliability.

**Nomenclature**

$\rho$ Density ($kg.m^{-3}$)
$C_p$ Heat capacity ($J.kg^{-1}.K^{-1}$)
$T$ Temperature (K)
$\kappa$ Thermal conductivity ($W.m^{-1}.K^{-1}$)
$Q_e$ Heat source ($W.m^{-3}$)
$\varphi_{rsa}$ Surface to ambient radiative heat flux ($W.m^{-2}$)
$\sigma_s$ Stefan Boltzmann constant (5.67E-8 $W.m^{-2}K^{-4}$)
$\epsilon$ Emissivity
$T_{air}$ Air temperature (K)
$\varphi_{csa}$ Convective heat flux ($W.m^{-2}$)
$h_{ia}$ Surface conductivity ($W.m^{-2}.K^{-1}$)
$J$ Surface radiosity ($W.m^{-2}$)
$G$ Irradiation flux ($W.m^{-2}$)
$n$ Refractive index
$e_b(T)$ Surface radiation produced ($W.m^{-2}$)

$\mu_r$ Complex relative permeability
$\varepsilon_r$ Complex relative permittivity
$\mu_r''$ Relative permeability imaginary part
$\varepsilon_r''$ Relative permittivity imaginary part
$k_0$ The vacuum wave number (rad.m$^{-1}$)
$\sigma$ The electric conductivity (S.m$^{-1}$)
$\varepsilon_0$ The vacuum permittivity (8.854187817...×10$^{-12}$ F.m$^{-1}$)
$\mu_0$ The vacuum permeability (1.2566370614...×10$^{-6}$ T.m/A)
$j$ The complex number
$\omega$ The angular frequency (rad.Hz)
$\boldsymbol{E}$ Electric field strength (V.m$^{-1}$)
$\boldsymbol{H}$ is the magnetic field intensity (A.m$^{-1}$)

## 1. INTRODUCTION

Microwave processing of materials is an emerging technology very appropriate for the synthesis or sintering of materials [1]. Recently, Kitchen et al. have reported different results related to the microwave synthesis of various types of materials, including oxides, carbides or nitrides [2]. In a more detailed way, Croquesel et al. have presented a specific instrumentation on a 2.45 GHz single-mode cavity for sintering alumina [3]. In terms of microwave-mater interactions, Mishra and Sharma have discussed in details how ceramic materials are heated by microwave depending on their physical properties [4]. In the same way, Oghbaei and Mirzaee [5] have summarized the different advantages and challenges to face when microwave energy is used for the sintering of materials. The main advantages of microwave heating are the high heating rates and the volumetric heating thanks to the large penetration depth of the microwave radiation in dielectric materials (~ a few centimeters). Therefore, the temperature field is expected to be homogeneous and the overall heating efficiency can be high. Even flash microwave sintering process can be envisaged using direct microwave heating as shown by Biesuz and Sglavo [6]. There are many experimental microwave systems for material sintering and most of them are designed by laboratories themselves, therefore the heat treatment conditions cannot be easily compared from one experiment to another. For instance, for low-loss dielectric materials such as alumina, spinel or insulating perovskite etc., researchers usually use susceptors materials to initiate the heating. These susceptors are made from highly microwave absorbing materials such as silicon carbide, zirconia, etc. The effect of the susceptor material properties on the heating behavior was discussed by Heuguet et al. [7]. At low temperature, susceptors are directly heated by microwave and they allow the low-loss dielectric material to be heated mostly by thermal radiation. As the dielectric losses of the material increase with temperature, there is a critical temperature above which the sample starts to heat by itself. The latter process is called Hybrid Heating process as the heat is provided by both thermal heat flux coming from the susceptor and by direct conversion of the microwave energy in the material. Different types of susceptors can be found in the literature, including rod-like arrangement (picket-fence arrangement), powdered susceptors or tubular susceptors. Bhattacharya et al. have described these different susceptor geometries [8]. Otherwise, these susceptors are often used in many different microwave applicators (single mode, rectangular, cylindrical, multimode, etc.) at different frequencies (5,8 GHz, 2.45 GHz, etc.). Most of the time, this is an empirical process that leads to the final geometry and assembly, based on considerations such as heating and temperature control; temperature distribution and rates, etc. In this work, our aim is to develop a fully controlled microwave heating process to sinter low-loss dielectric refractory materials. For this purpose, a 915 MHz single-mode rectangular cavity was chosen as it provides a high volume chamber and, compared to the common and widely spread 2.45 GHz frequency, the 915 MHz one offers a larger penetration depth as discussed by Zhai [9]. In this work, an assembly consisting of a thermal insulation material, equipped with two SiC succeptors plates, was used to sinter large size alumina in a 915 MHz cavity. Our wish was to provide a hybrid heating process, with very good temperature regulation and distribution. By doing so, a volumetric heating process is expected thanks to the well-established dielectric heating mechanism. Moreover, the use of susceptors also allows the sample not to be too rapidly cooled from its surface, as a result, a homogeneous temperature distribution can be obtained. Modeling results will be also advantageously used to improve our understanding of the overall process, and to clearly point out the strong effect of the assembly geometry to the heating behavior. A fully automatization of the system will also be used based on our previous development [10] to sinter low dielectric losses alumina materials. The microstructure of the materials being sintered will be fully characterized and discussed with respect to the experimental conditions. A reliable process leading to a reproducible microstructure will be researched. Overall, this paper presents the strategy and the different results obtained on hexagonal shaped large alumina samples (>30 $cm^3$) fully sintered in our microwave rectangular original microwave cavity.

## 2. EXPERIMENTAL

### *2.1 Experimental 915 MHz Equipment*

Marinel et al. have previously described the 915 MHz microwave equipment [10]. Basically, the equipment consists of a 915 MHz source (SAIREM GLP 50KSM) which delivers a microwave power to a rectangular wave-guide. At the end of the microwave line, a single-mode applicator is used to heat up the material. The system is equipped with a 4 stubs automatic impedance plunger (AI4S from SAIREM) and a motorized short circuit piston/termination. A fully automatization was implemented allowing to program thermal cycle (Figure 1).

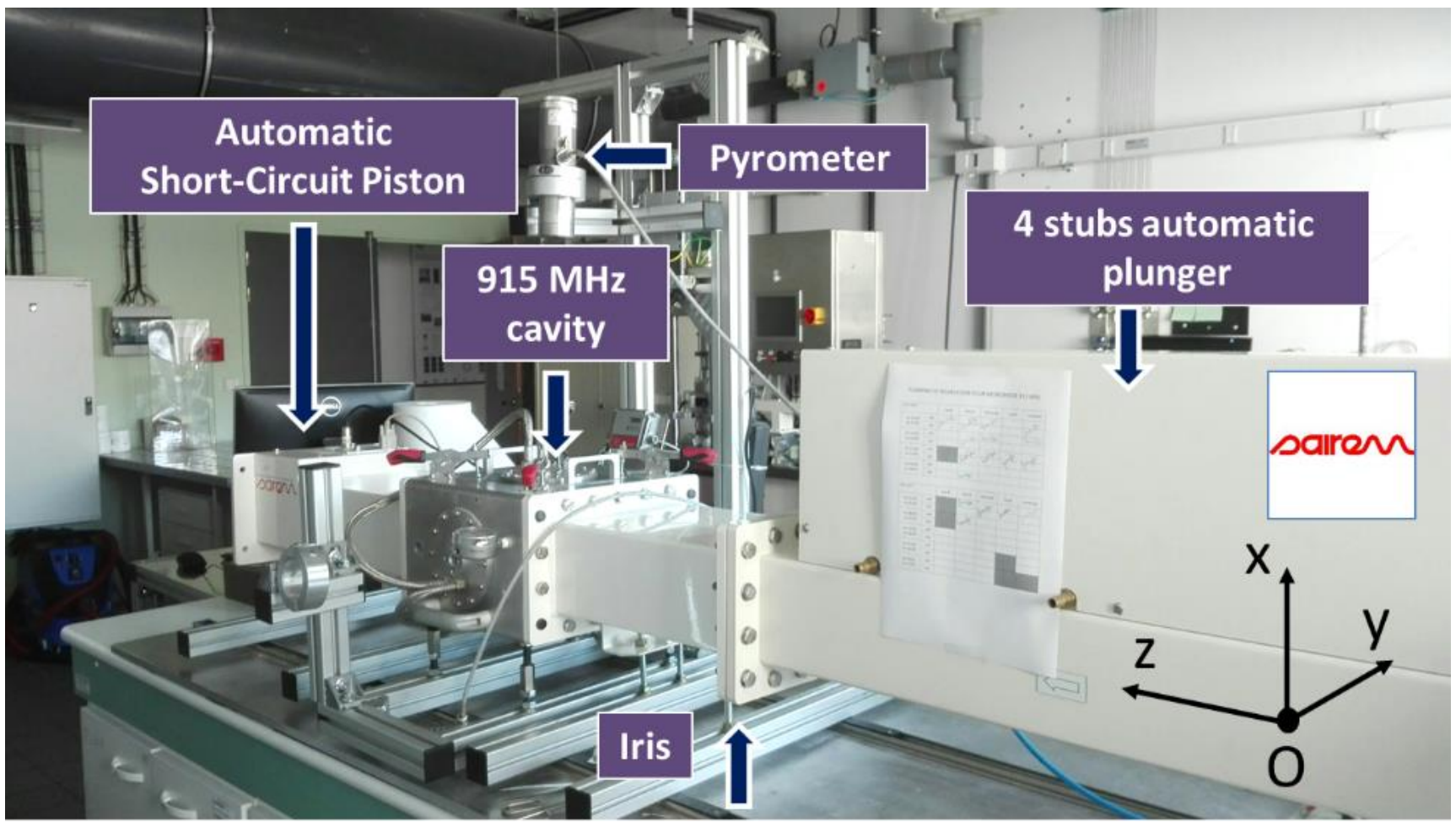

**Figure 1.** Photograph of the microwave system with the 4 stubs automatic impedance tuning device and the short-circuit piston.

The automatic system described elsewhere [10] mainly consists of using three internal and independent loops: one for maintaining the impedance matching, another one for keeping satisfactory the resonance condition and the last one for adjusting the incident microwave power through an auto-adaptive PID program. In order to tune the P, I and D values required to calculate the incident power value (incident power feedback loop), an initial power of 700 W is systematically applied at the beginning of the cycle, and the response of the system is measured (i.e. the temperature). This is shown on the figure 7 (see the red circle) where it can be seen the temperature response to the 700 W imposed incident microwave power. When the temperature achieves 460 °C, the power is automatically switched off and the cooling temperature is recorded. From the temperature response to the square signal, the P,I and D coefficients initial values are calculated. Then the temperature regulation is automatically working by continuously adjusting the incident microwave power, calculated from the automatic calculated P, I and D values. Therefore, the temperature signal can rigorously follow the programmed cycle.

### *2.2 Numerical Analysis*

Understanding previous described microwave system requires solving numerically a very complex Multiphysics problem where the microwave physics, resonance phenomena and thermal dependence of material properties are key parameters as described by Maniere et al. [11,12]. These parameters are coupled to each other as the heating of the specimen and the tooling modifies the microwave properties which in turn influence the cavity microwave distribution and the resonance performances. Maniere et al. have shown that the PID regulation system has to deal with the microwave cavity response to dimensional changes and with the temperatures change [13]. To show the interconnection between the microwave fields distribution, the heating and the stability of the thermal distribution, a Multiphysics simulation is necessary. This numerical tool is based on the finite element method (FEM). It encompasses the physics of microwaves through Maxwell's equations and the heat transfer through the heat equation and surface-to-surface thermal irradiation between the heating tool/specimen. This model combining the electromagnetism and the heat transfer equation was previously implemented by Maniere et al. [11]. Convection is another heat transfer mechanism which is classically simulated by fluid dynamic physics. Both Egorov [14,15] have already introduced the contribution of the convection phenomenon into their numerical model. However, due to the small distances between the susceptors and the specimen (few mm), the convection can be neglected here. Comsol Multiphysics is the FEM code used in this work. In this code, Maxwell's equations are combined to describe the microwave distribution using the following equation:

$$\nabla \times (\mu_r^{-1} \nabla \times \boldsymbol{E_r}) = k_0^2 \left( \varepsilon_r - \frac{j\sigma}{\omega\, \varepsilon_0} \right) \boldsymbol{E_r} \tag{1}$$

with $\boldsymbol{E_r}$ defined by the harmonic electric field expression $\boldsymbol{E} = \boldsymbol{E_r} exp(j\omega t)$.

The heat transfer part of the electromagnetic-thermal model is governed by:

$$\rho C_p \frac{\partial T}{\partial t} + \nabla.(-\kappa \nabla T) = Q_e \tag{2}$$

with dissipated microwave source term:

$$Q_e = \frac{1}{2}(\varepsilon_0 \varepsilon_r'' \boldsymbol{E}^2 + \mu_0 \mu_r'' \boldsymbol{H}^2) \tag{3}$$

The microwave boundary conditions are a rectangular TE10 port which simulates the microwave incident power and the absorption of the reflected power (role of the waveguide circulator in the experimental setup). The other external surfaces are assumed perfectly conductive, reflecting and confining the waves inside the cavity. The thermal problem is limited to the insulation box, the susceptor and the specimen. The external insulation box boundary conditions are convection and radiation using:

$$\varphi_{csa} = h_{ia}(T_{air} - T) \tag{4}$$

$$\varphi_{rsa} = \sigma_s \epsilon ({T_{air}}^4 - T^4) \tag{5}$$

with the coefficients values $h_{ia} = 5\, Wm^{-2}K^{-1}$ and $\epsilon = 0.83$. The internal surface to surface thermal radiation is calculated considering each point emitted $e_b(T)$ and incoming radiation G. The relation between these quantities is defined through the radiosity $J$ expression which is the total outgoing thermal radiative flux:

$$J = (1 - \epsilon)G + \epsilon e_b(T) = (1 - \epsilon)G + \epsilon n^2 \sigma_s T^4 \tag{6}$$

As this study focuses on the microwave heating, the sintering is not simulated. All the temperature dependent properties used in the simulation are summarized in the Appendix table A [13].

*2.3 Materials Preparation and Characterization and the microwave assembly*

Alumina from Baïkowski was selected ($Al_2O_3$, BMA15, 99.99 %, D50=150 nm, 16m$^2$/g) and large hexagonal shaped samples were prepared by SOLCERA by slip casting (59 mm ×10 mm thickness, see figure 2). A green density of 60 % was obtained (theoretical density of alumina=3,97 g/cm$^3$). Samples were conventionally sintered in air in our 915 MHz single-mode applicator. Samples were subsequently polished to a final step with colloidal silica suspension (STRUERS OP-U Non-Dry) and SEM microstructures were collected with ZEISS Supra 55. Density of materials was measured using Archimedes' method in ethanol and grain sizes were measured using the intercept method. Finally, the hardness was measured using Vickers indentation technique at 1 kgF for 15 seconds (CLEMEX MMT with digital camera).

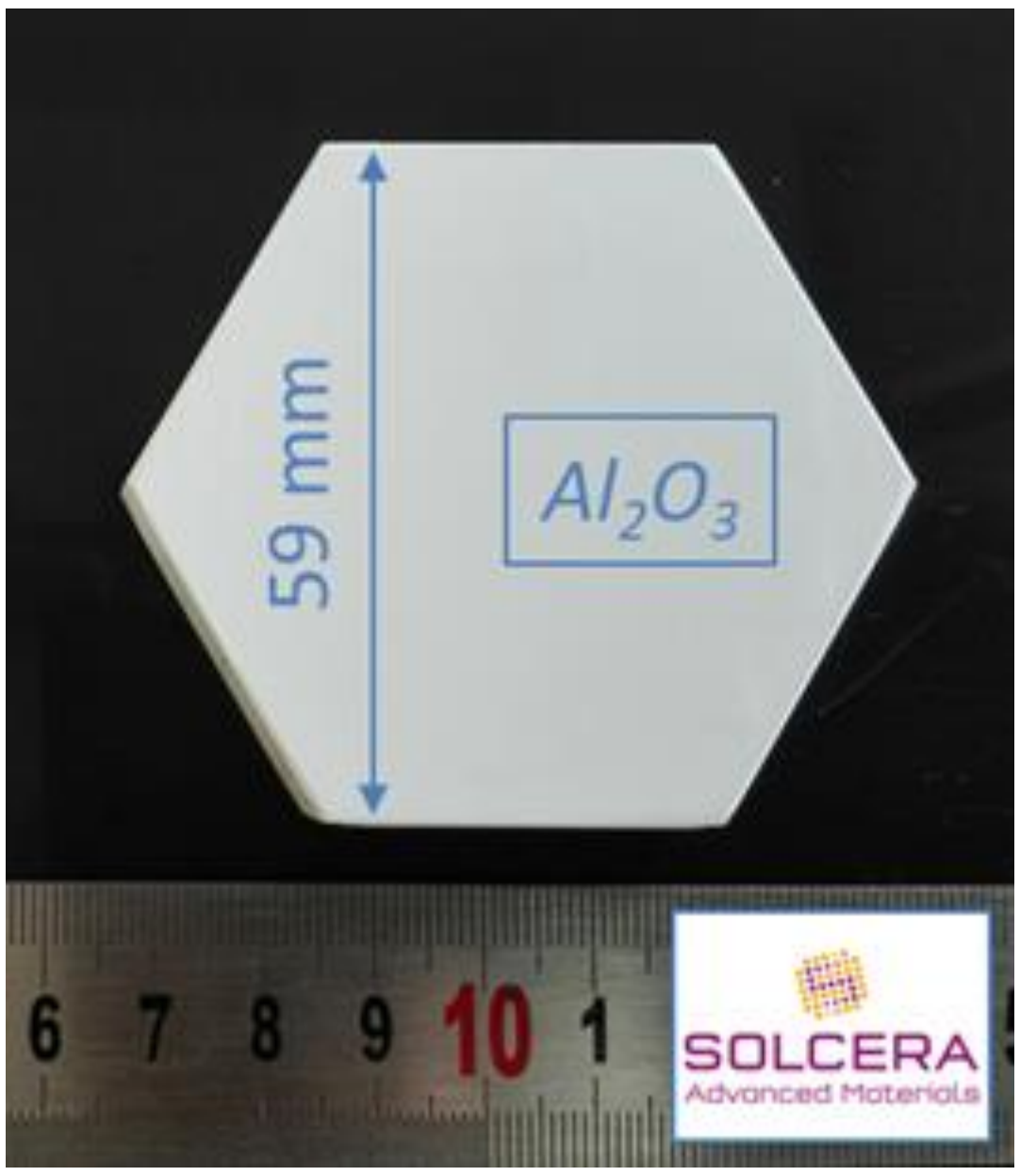

**Figure 2.** Picture showing a green sample of hexagonal shaped alumina material. Samples were shaped by slip casting by SOLCERA Advanced Materials company.

The microwave assembly designed for our purpose is shown on figure 3. The alumina hexagonal sample is vertically placed onto an alumina support piece and the larger side of the sample is parallel to the z direction (see figure 1). SiC plates (80 mm×80 mm×7mm, Anderman Ceramics AC-RSiC-A, Recrystallized Silicon Carbide) are placed parallel to the sample in between as schematically shown on figure 3. This design was found to be optimal for homogenizing the temperature field within the material. Otherwise, as seen further, the modeling data will support this original geometry. A thermal insulation box is machined from an alumina silicate porous foam from RATH© and allows to thermally insulate the assembly. The TE 105 mode is tuned therefore the sample is located where the electrical field is maximum. A single-color pyrometer (IRCON modline 5 G) with a temperature range of 350°C<θ<2000°C is used for measuring the temperature, targeting on the upper side of the sample. Macaigne et al. [16] measured an apparent emissivity of ~ 0,7 at about 1500°C on alumina. In this work, a similar apparent emissivity was found and this value was selected for all experiments. Practically, temperature calibration consists in measuring the 'apparent emissivity' of the sample. For that, a piece of calibrating material is placed on a tiny hole in the sample surface and when the calibrating material starts melting, the emissivity of the IR pyrometer is tuned so that the measured temperature equals the calibrating material melting point (a CCD camera is used to observe the sample surface with the calibrating material). It is assumed that the calibrating material is at the same temperature (or very close) to the sample surface temperature. $Nb_2O_5$ oxide (melting point~ 1512°C) material was used as calibrating material. By doing so, an apparent emissivity of 0.7 was found; more details about this calibration can be found in reference [16].

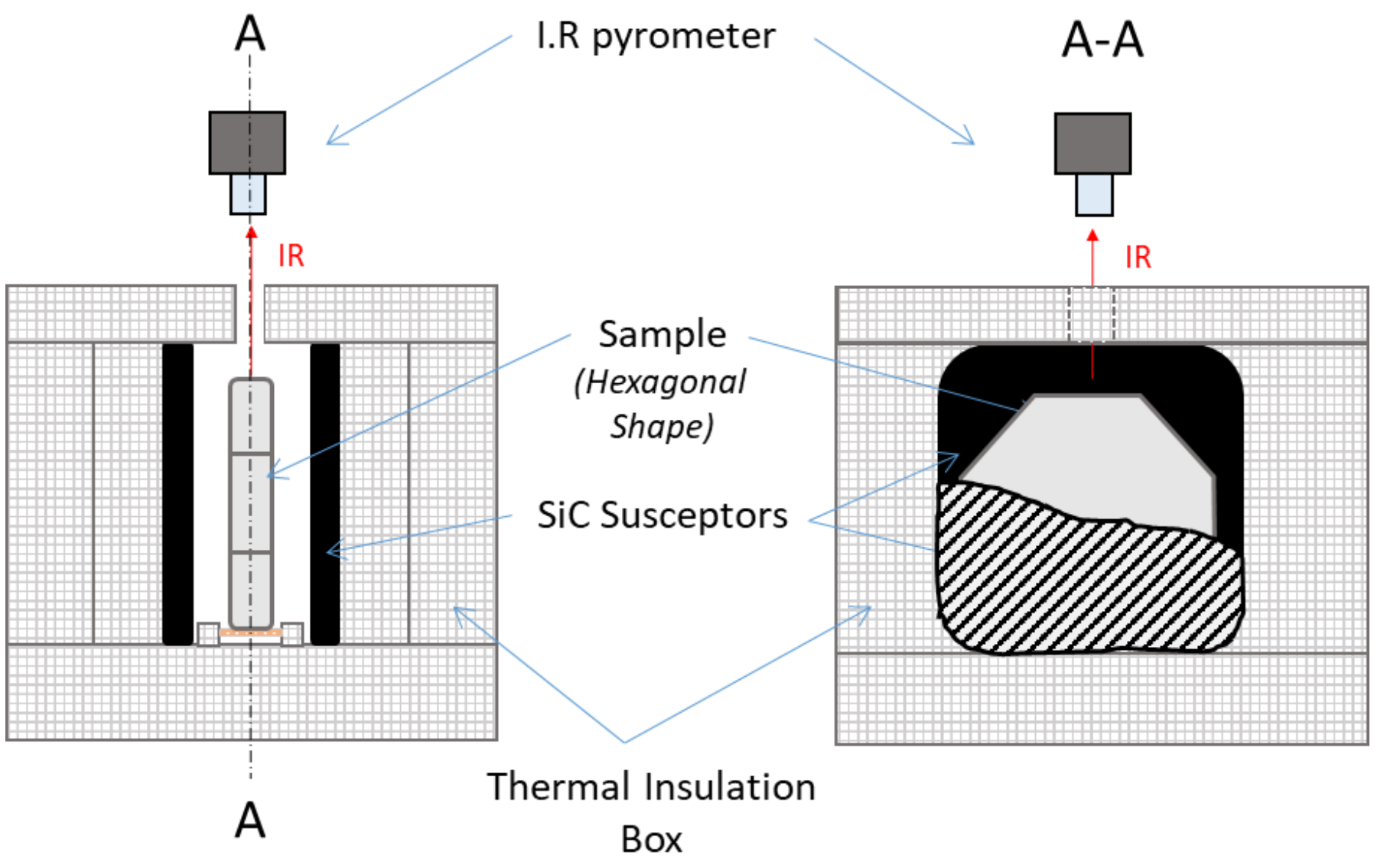

**Figure 3.** Assembly used to heat up the green ceramic sample into the 915 MHz single-mode cavity.

## 3. RESULTS AND DISCUSSIONS

In this section, the microwave cavity is first presented through a simulation approach. The microwave distribution and heating are discussed. Then, the experimental results obtained for the hexagonal shaped samples are presented and discussed with respect to simulation.

### *3.1 Microwave Cavity*

The heating strategy of the 915 MHz cavity is first presented through pure microwave calculation. These steady state parametric simulations consider a TE10 915 MHz microwave port, the iris, thermal insulation, SiC susceptors plates and alumina hexagons. The main parameters are studied in order to explore this cavity potential, the orientation of the parallel SiC plates vs E field, the distance between the iris and the specimen area, and, for each case, the total length of the cavity. The last parameter is essential as it allows locating the resonance (pseudo-resonance) phenomena which are highly sensible to the cavity dimensions at a given frequency. The microwave port parameter "S11" which quantifies the amount of reflected microwave (reflection coefficient) power is used to locate the resonance phenomena, as it was previously described by Yakovlev et al. [17]. Resonance is associated with a minimal value of S11. In figure 4 is presented the S11 curves and E field distribution for three SiC plate orientations. This shows clearly the strong influence of the susceptor orientation with respect to the electrical field. When the susceptors plates are perpendicular to the electrical field, the coupling with microwaves is low and the resonant profile is very sharp, with classically a very high E field when close to the resonance. This abrupt resonant profile allows obtaining high field intensity but those conditions are very difficult to control due to the very narrow resonance peak. On the other hand, when the susceptors plates are oriented parallel to the electrical field, the resonant profile is nearly annihilated (highly conductive SiC) and this pseudo-resonant profile shows a very large peak with a small intensity shifted towards lower cavity length value (peak position is going from ~ 485 mm down to ~ 440 mm). In this configuration the heat dissipation in the susceptor is expected to be optimal, and more homogeneous within the material. Such a configuration is preferred for an easy regulation control, with a wide and not too deep resonance profile.

In order to find the optimal position of the iris to locate the specimen in a maximal electrical field, another parametric study has been investigated where the iris position and cavity length are explored. The results are reported in figure 5. The iris position has an impact on the intensity and location of the pseudo-resonant profile. The optimal configuration in terms of the electric field intensity and position corresponds to the case where the iris is moved 60 mm closer to the specimen from initial position. In summary, the strategy adopted in this microwave furnace consists to avoid too abrupt and intense resonant mode and replace it with a pseudo-resonant mode which is more stable in a configuration where the microwave coupling with the susceptors is optimal.

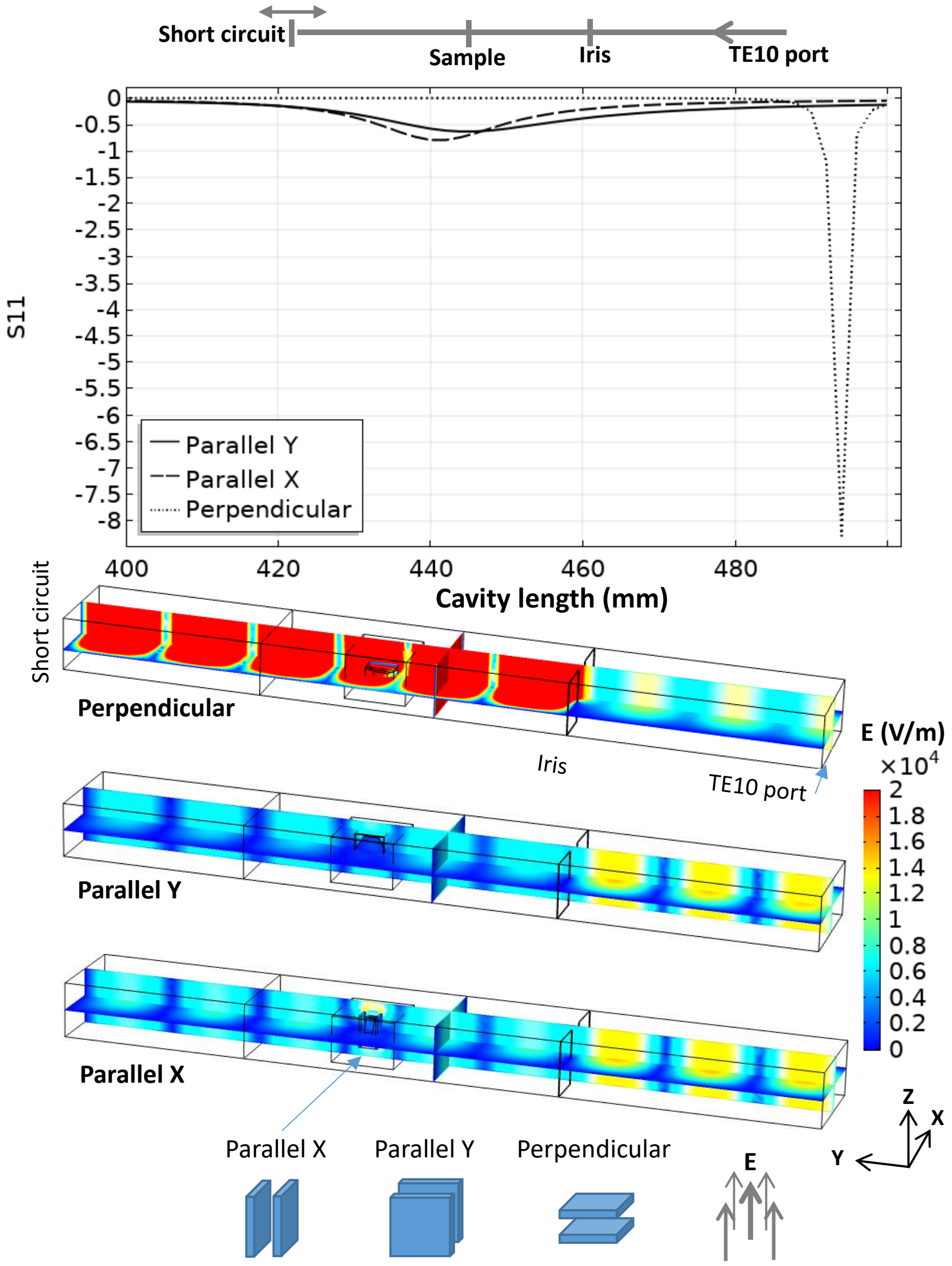

**Figure 4.** Resonant behavior of the 915 MHz cavity for three different orientations of the SiC susceptors plates with the E field; the simulation image corresponds to the minimum S11 parameter for each tested orientation.

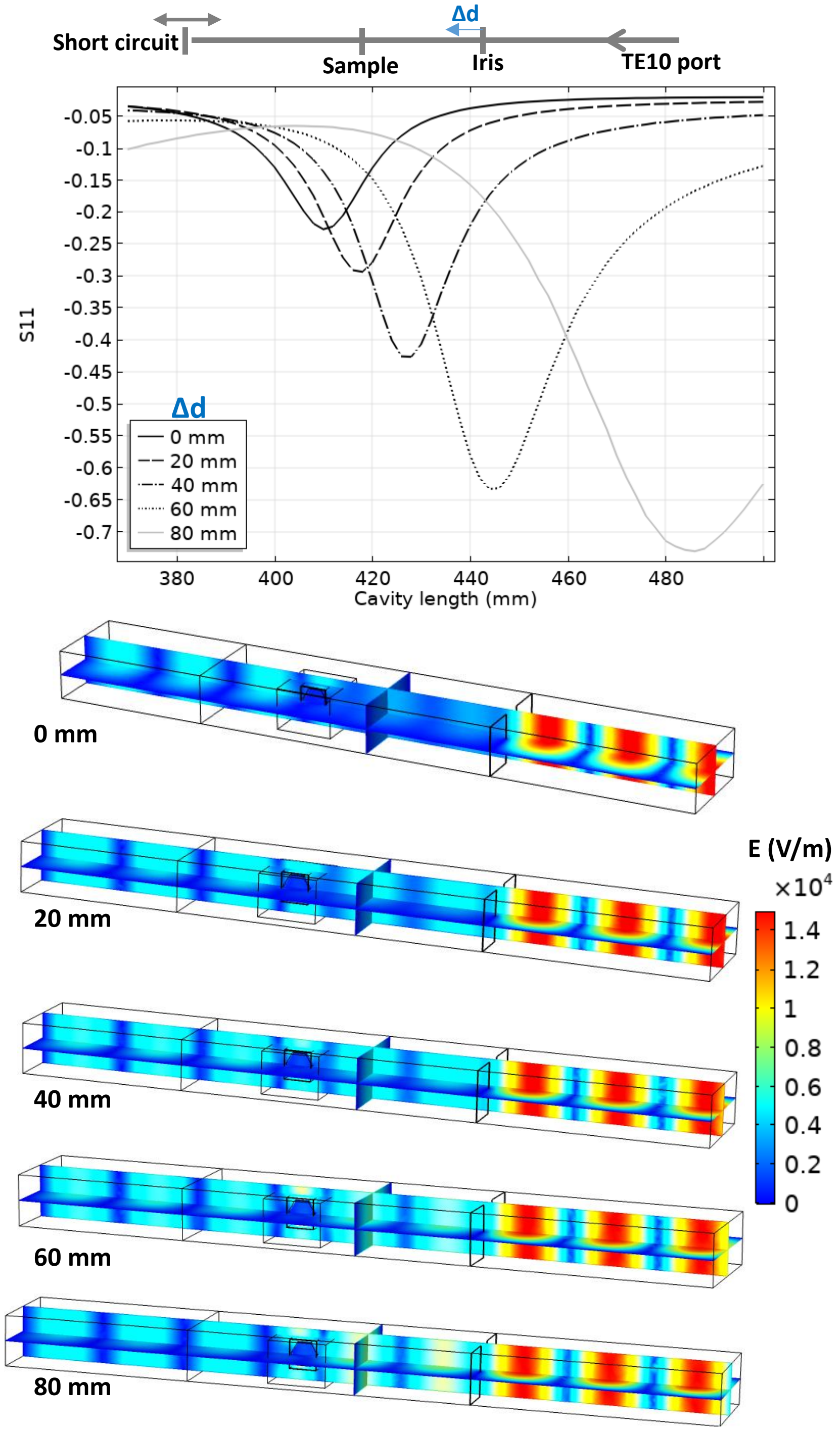

**Figure 5.** Resonant behavior of the 915 MHz cavity for different iris positions; the simulation images corresponds to the minimum S11 parameter for each tested position.

### *3.2 Simulated Microwave Heating*

The microwave heating is simulated in the "parallel Y" configuration (see figure 4) for a maximum microwave coupling of the SiC plates and in an iris configuration where the electric field is centered on the sample ("60 mm configuration", see figure 5). The simulated temperature curves on the sample/susceptor and the thermal/electrical field during the heating, holding and cooling are reported in figure 6. The microwave incident powers were digitally PID regulated to impose the thermal cycle which will be used experimentally (see next section). Because the heating of the alumina specimen is highly delayed compared to the susceptor (high thermal response), the PID regulated temperature is the average temperature between the susceptor and the sample (see figure 6). This way highly stabilizes the temperature regulation and is closer to the experimental conditions where the thermal radiation coming from the susceptor participates in the apparent temperature measured by the pyrometer which is pointed on the alumina sample upper surface, as discussed by Croquesel et al. [3]. As shown in figure 6, the temperature difference between the susceptor and the sample is about 500 °C at the beginning and this difference is decreasing down to a few degrees at 1000°C. After this temperature, the temperature distribution is very homogeneous between the two susceptor plates. The analysis of the heat flux (at high temperature) in the sample surface shows that the heat is transmitted to the sample mostly by thermal radiation from the susceptor to the sample. The dominant thermal exchange by radiation can be explained by the $T^4$ law (5). Because the alumina has a low dissipation factor compared to the susceptor (SiC), the heating of the alumina hexagon is stabilized by the high external heat coming from the susceptor. Compared to the direct microwave heating which is known to have high inherent heating instability such as hot spots, the present study provides heating as stable as conventional sintering. Maniere et al. have clearly showed by modeling the advantage of using an hybrid configuration for stabilizing the thermal process [18]. Because the heating elements (susceptor) are closed to the specimen, faster heating cycles are allowed. It can be noted that despite a low microwave-alumina coupling, a high electric field (2 kV/m) is present in the sample volume. This high electric field may enhance the grain boundary diffusion, as suggested by Olevsky *et al.* [19]. This effect may accelerate the sintering kinetics under microwaves as proposed by Croquesel [3] and Olevsky et al. [19]

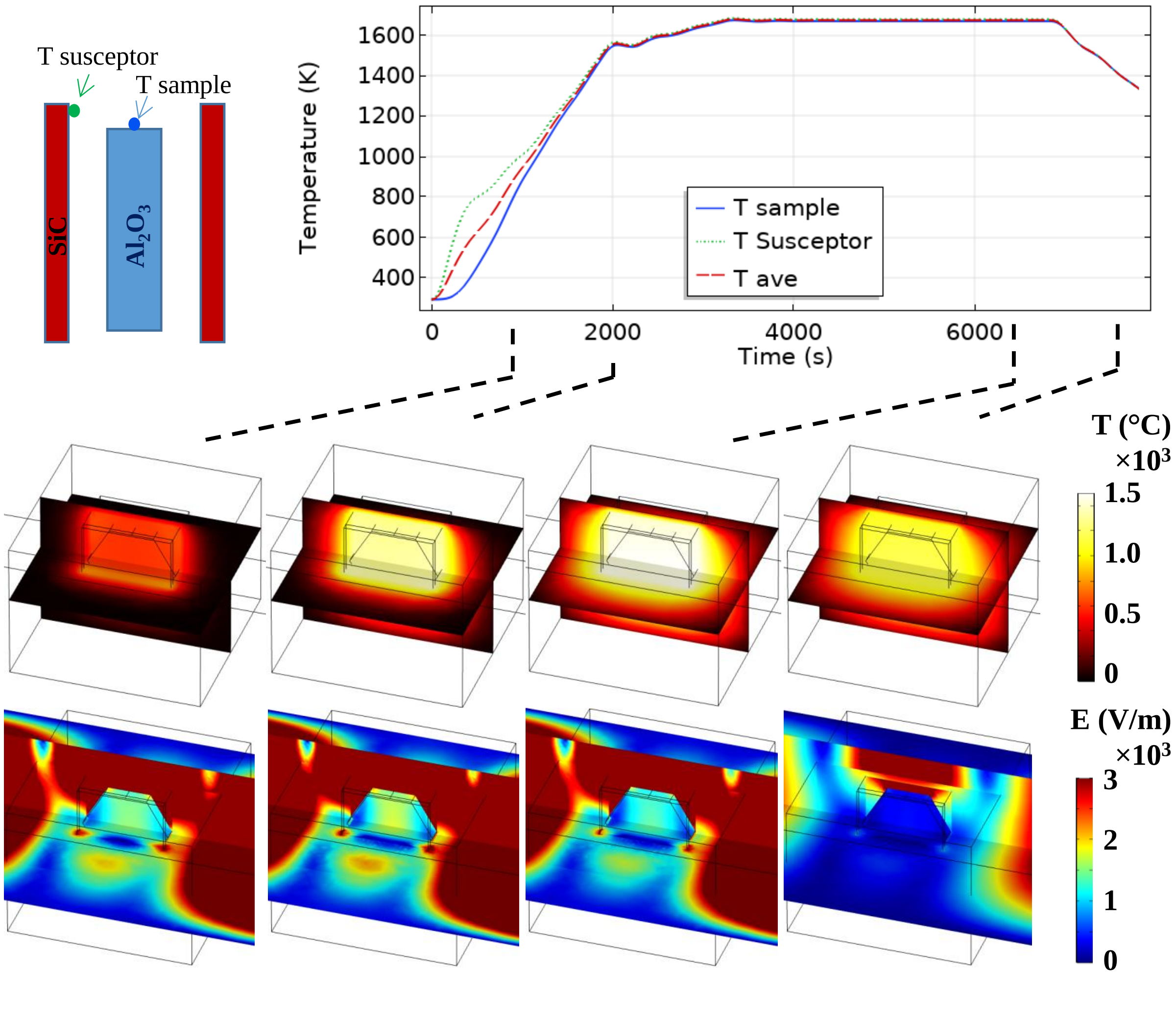

**Figure 6.** Electromagnetic-Thermal simulation of microwave heated alumina specimen.

*3.3 Temperature Cycle and Sample Density*

The geometry optimized by the simulation study was used to sinter hexagonal alumina samples in our 915 MHz cavity. The chosen temperature cycle is shown in figure 7 and was defined from previous experiments on alumina. Two heating ramps were programmed: one at 27°C/min from RT to 1250°C (1523 K to compare to modeling) and the second one, at 7 °C/min, from 1250°C to 1400°C (1673 K). It was indeed necessary to slow down the heating ramp at high temperature as the thermal losses (especially from thermal radiation) are very high at elevated temperature. Therefore, the system could not follow the fastest imposed heating ramp (of 27°C/min). The dwell temperature of 1400 °C (1673 K previously modeled) was found to be appropriate to get high density while avoiding significant grain growth. The sample density measured on the whole hexagonal sample is 99,6% of the theoretical value. The density was also measured on the 6 different triangles, previously cut from the hexagon. As shown figure 7 (see inset), it can be seen that the resulting density is similar from one location to another, showing that the sintering process is homogeneous. In order to check the reproducibility of the microwave heating system, several hexagons were sintered by using exactly the same conditions and thermal cycle. After sintering, all samples had the same density of ~ 99,6 % of the theoretical value. As illustration, the figure 8 shows an assembly of seven samples before and after sintering and clearly proves the good reproducibility of the as-developed microwave process.

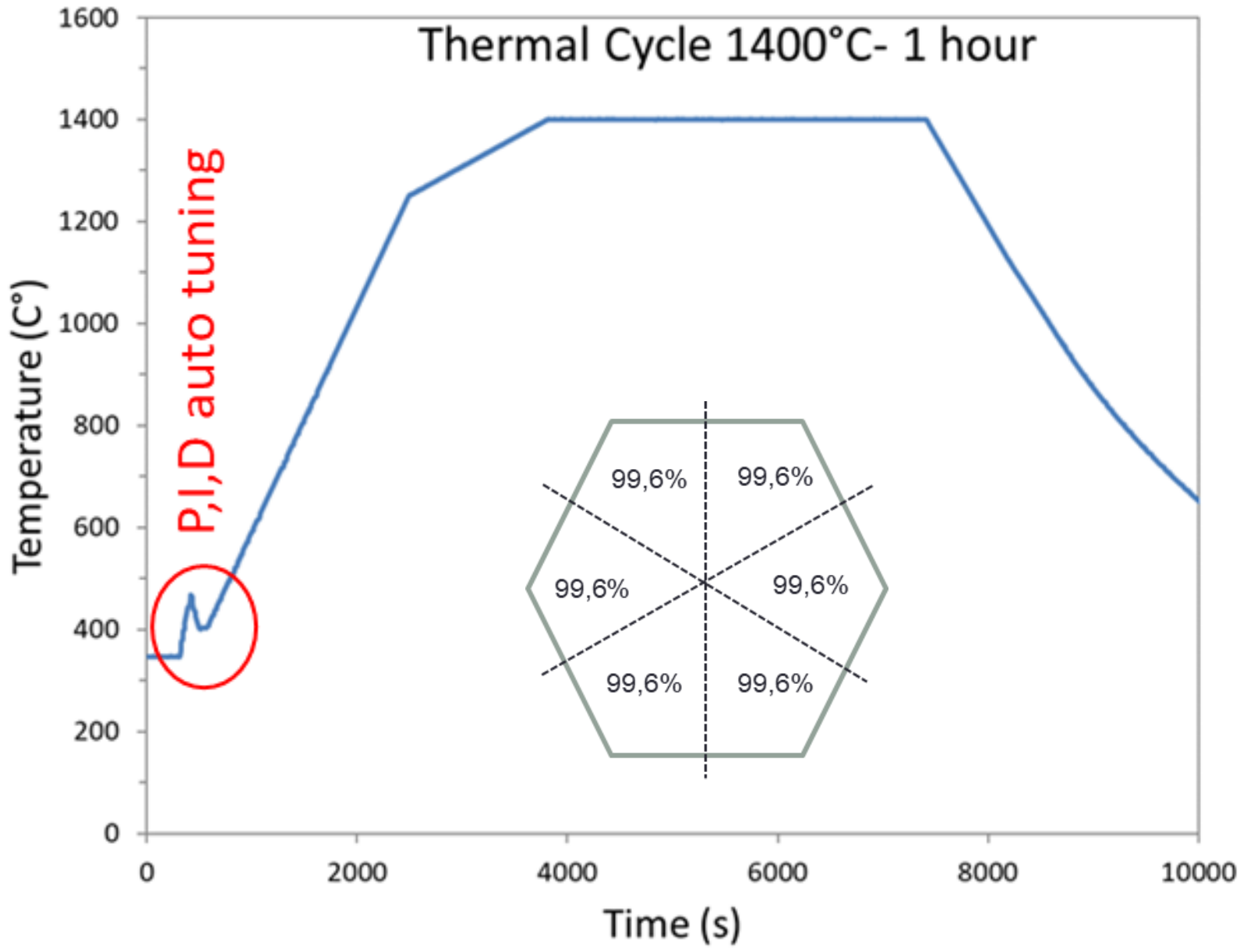

**Figure 7.** Heating program cycle imposed to sample during microwave sintering in our 915 MHz single-mode cavity.

The figure 9 shows the typical SEM microstructures of the alumina samples at different locations. First of all, the microstructures are dense and it can be clearly seen that the grain size is slightly smaller near the surface than in the bulk. The grain size goes from around 1 µm at the sample surface to about 1.9 µm in the center. This shows that despite the use of the SiC susceptors and its significant contribution in the heating (see the modeling section), the center of the material does not densify with delay compared to its periphery. This is consistent with a hybrid heating mode, which results from the large microwave penetration depth into the material (at 915 MHz, 1400°C, the penetration depth into alumina, calculated from the data table A, gives a value of 247 cm) and the quite high microwave electrical field (>2kV/cm) which develops within the bulk alumina material, as estimated by modeling.

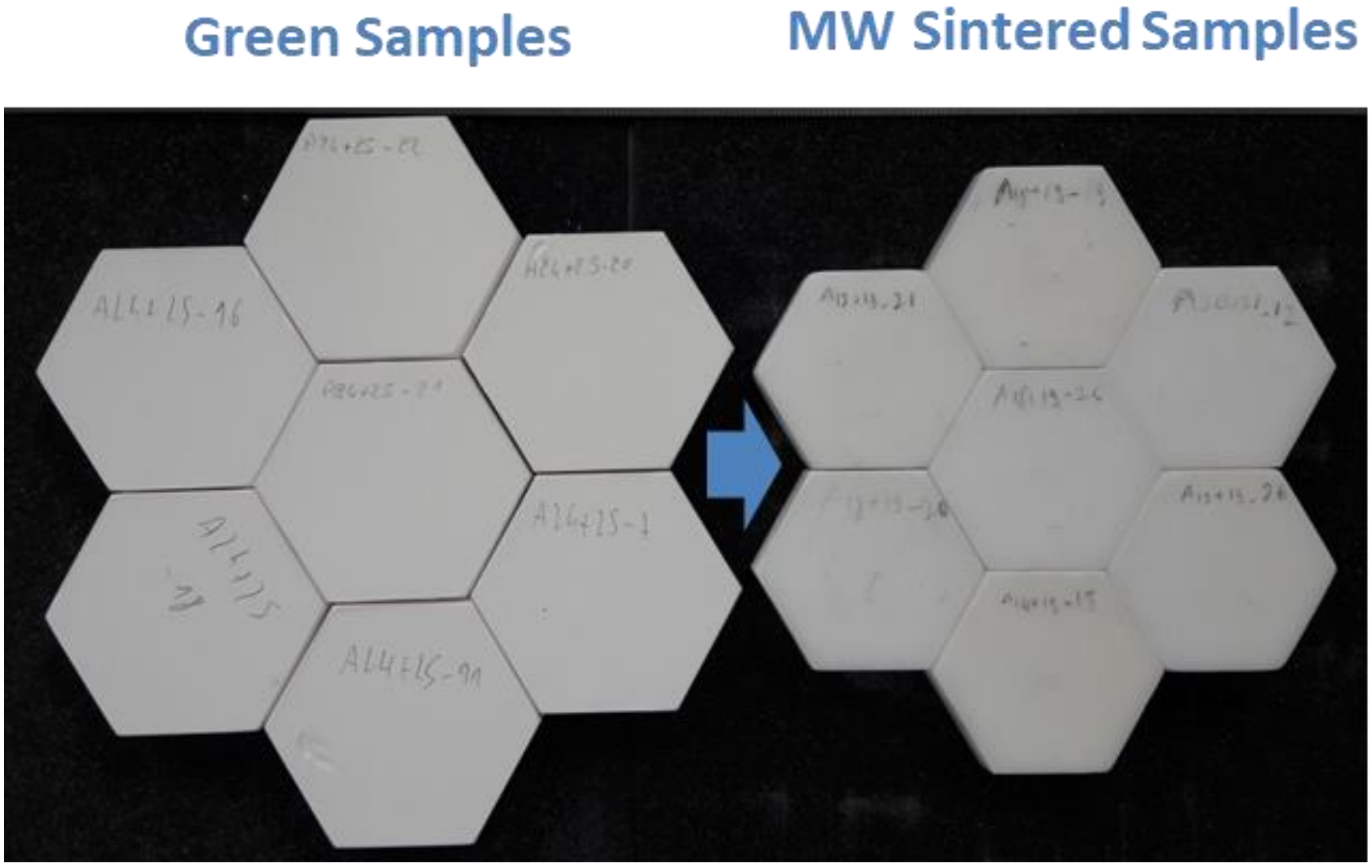

**Figure 8.** Several hexagon alumina samples microwave sintered.

This grain size evolution is also reflected by the hardness value dependence with the position through the sample cross section of (figure 10). The hardness value is significantly higher in the sample periphery than in the bulk. An eye guide is plotted in red in the graph to ease the visualization of the hardness dependence with position. This hardness dependence is well correlated to the grain size distribution observed by SEM. Otherwise, the hardness value obtained is very close the very high hardness value measured on submicronic grain size sintered samples by spark plasma sintering process on the same alumina powder (up to HV1~21 GPa)[20].

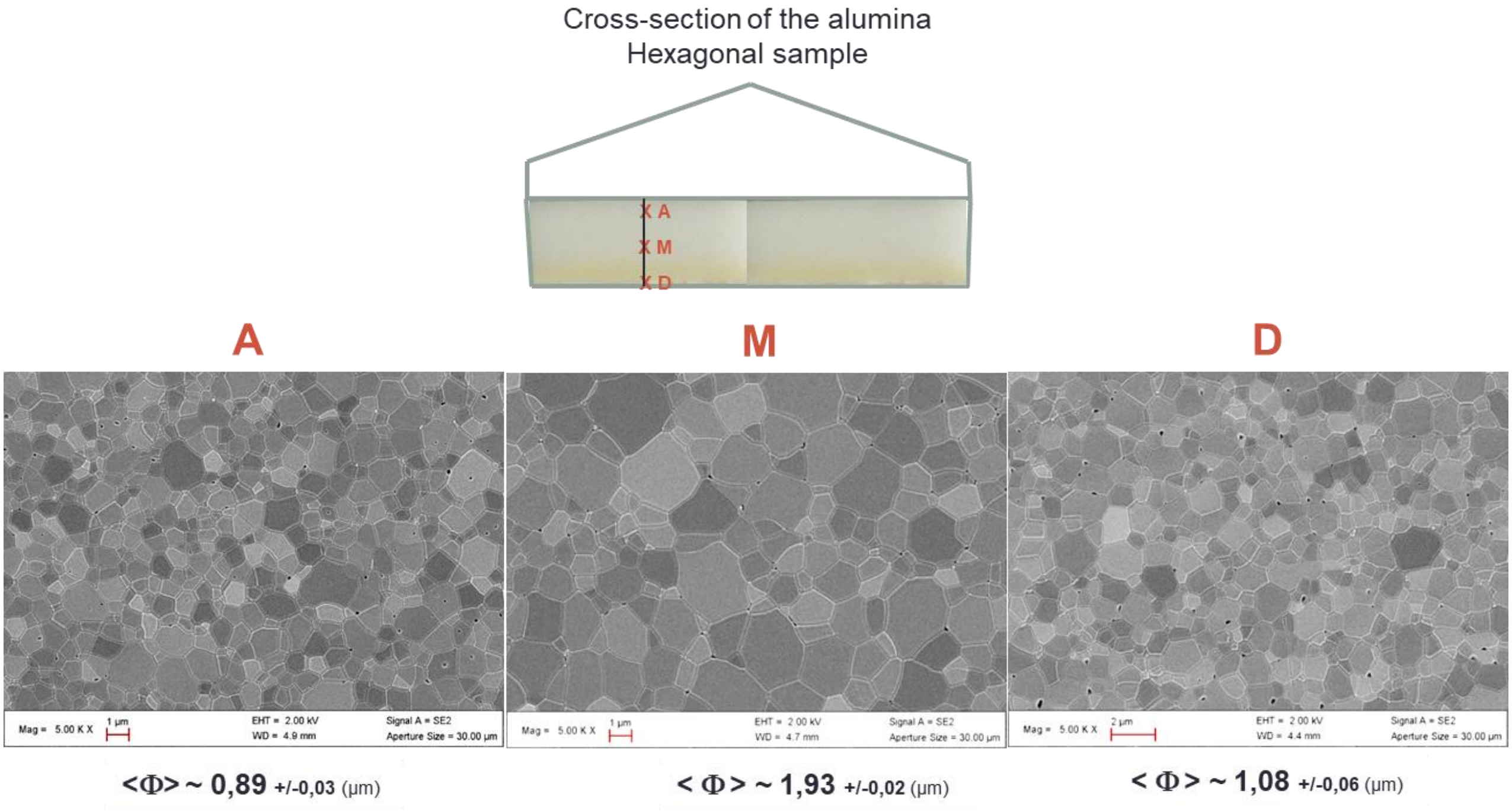

**Figure 9.** Typical SEM microstructures collected at the sample surface (A and D) and in the middle of the alumina hexagonal microwave sintered materials.

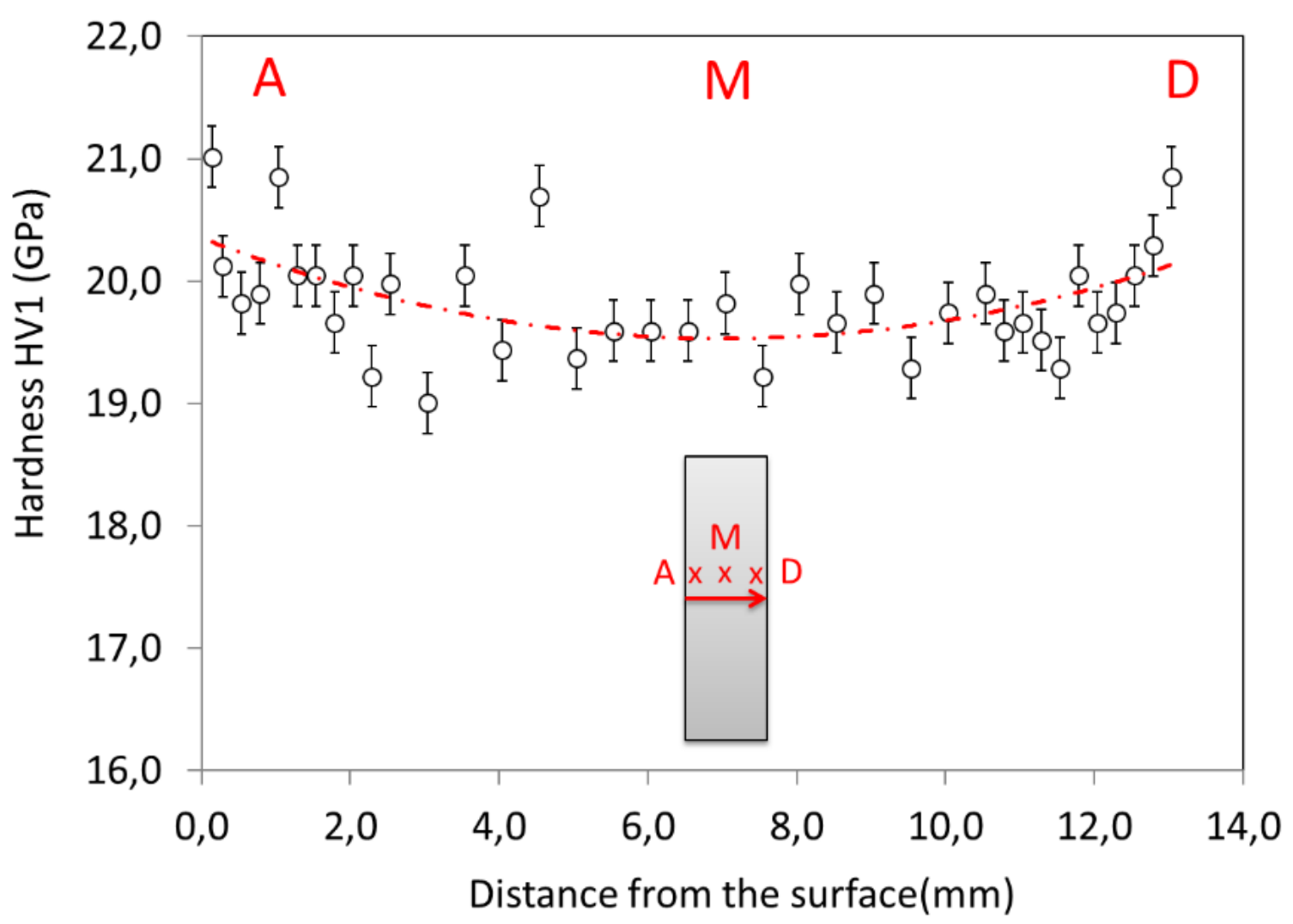

**Figure 10.** Plot of the hardness value as a function of the distance from the surface on a sample cross-section.

## 4. CONCLUSION

This work aimed at developing a fully automatized 915 MHz single-mode cavity to sinter large alumina samples. Modeling, temperature regulation, reproducibility of the process and materials characterizations were implemented and investigated. Through this research, an original susceptors SiC-plates assisted microwave process was designed in a rectangular 915 MHz microwave cavity. From the modeling results, it was shown that the most appropriate geometry was to align the largest surface of the SiC susceptors parallel to the electrical field. In the same way, the two plates susceptors were aligned along the microwave propagation direction. By doing so, a wide and not too high resonance mode was applied, in resulting a sample heating homogeneous with a very efficient temperature regulation process. Otherwise, this system was demonstrated to provide a hybrid heating mode as shown by modeling. Experimentally, the grain size distribution reveals that grain size at the center is slightly larger than in the sample periphery, which could also be consistent with a hybrid heating mode. In terms of mechanical properties, the materials characterization reveals that highly dense materials (>99,6% dense) were produced exhibiting high hardness values (HV1~ 20 GPa). Finally, the as-developed process was found to be reproducible clearly showing that these developments could bridge the gap in the use of microwave technology for high temperature sintering applications.

## Appendix A

**Table A: Electromagnetic-thermal properties of the insulating box and silicon carbide susceptor and alumina (T is temperature in K).**

| Material | | Temperature range (K) | Expression |
|---|---|---|---|
| Insulation | *Cp* (J .kg$^{-1}$.K$^{-1}$) | 273-1600 | -5.31E-4T$^{2}$+1.25T+5.18E2 |
| | *κ* (W.m$^{-1}$.K$^{-1}$) | 273-1700 | 1.70E-2+1.4E-4T |
| | *ρ* (kg .m$^{-3}$) | 273-1700 | 4.43E2-1.04E-2T |
| SiC | *Cp* (J .kg$^{-1}$.K$^{-1}$) | 273-673<br>673-1573<br>1573-1700 | -8.35+3.08T-0.00293 T$^{2}$+1.0268E-6 T$^{3}$<br>772+0.431 T-2.10E-5 T$^{2}$<br>1400 |
| | *κ* | 273-1700 | 192-0.326 T+2.74E-4 T$^{2}$-7.71E-8 T$^{3}$ |
| | (W.m$^{-1}$.K$^{-1}$) | | |
| | $\rho$ (kg .m$^{-3}$) | 273-1700 | 2977+0.0510 T-2.29E-4 T$^2$+2.98E-7 T$^3$-1.92E-10 T$^4$+4.77E-14 T$^5$ |
| $Al_2O_3$ | $Cp$ (J .kg$^{-1}$.K$^{-1}$) | 273-1700 | 850 |
| | $\kappa$ (W.m$^{-1}$.K$^{-1}$) | 273-1700 | 39500T$^{-1.26}$ |
| | $\rho$ (kg .m$^{-3}$) | 273-1700 | 3899 |
| mAir | $Cp$ (J .kg$^{-1}$.K$^{-1}$) | 273-1600 | 0.177T + 961 |
| | $\kappa$ (W.m$^{-1}$.K$^{-1}$) | 273-1600 | 6E-05T + 0.0035 |
| | $\rho$ (kg .m$^{-3}$) | 273-1700 | 0.02897P/(RT) |
| | Dynamic viscosity (Pa s) | 273-1700 | 2.82E-6 + 7.51E-8T-3.01E-11T$^2$ + 8.88E-1T$^3$ -1.01E-18T$^4$ |
| Insulation<br>SiC<br>$Al_2O_3$ | Emissivity ϵ | 273-1700 | 0.83<br>0.9<br>0.8 |
| SiC | $\varepsilon_r'$ | 273-1700 | 5.30E-7T$^3$-8.18E-4T$^2$+4.13E-1T+3.00 |
| | $\varepsilon_r''$ | 273-1700 | 3.60E-7T$^3$-5.06E-4T$^2$+2.52E-1T+55.2 |
| Insulation | $\varepsilon_r'$ | 273-1700 | 5.03E-8T$^2$+1.37E-5T+1.5 |
| | $\varepsilon_r''$ | 273-1700 | 2.5E-9T$^2$-1.64E-6T+3.96E-4 |
| $Al_2O_3$ | $\varepsilon_r'$ | 273-1700 | 1.34E-3T+8.3 |
| | $\varepsilon_r''$ | 273-1700 | 4.62E-5T-8.87E-3 |